 \documentclass[pdflatex,sn-mathphys-num]{sn-jnl}
 
\usepackage{graphicx}%
\usepackage{multirow}%
\usepackage{amsmath,amssymb,amsfonts}%
\usepackage{amsthm}%
\usepackage{mathrsfs}%
\usepackage[title]{appendix}%
\usepackage{xcolor}%
\usepackage{textcomp}%
\usepackage{manyfoot}%
\usepackage{booktabs}%
\usepackage{algorithm}%
\usepackage{algorithmicx}%
\usepackage{algpseudocode}%
\usepackage{listings}%

\usepackage{tabularx}
\theoremstyle{thmstyleone}%
\theoremstyle{thmstyletwo}%

\theoremstyle{thmstylethree}%

\begin{document}

\title[Two-Dimensional Structural Characterization of Music Genre Communities in Playlist Co-occurrence Networks]{Two-Dimensional Structural Characterization of Music Genre Communities in Playlist Co-occurrence Networks}

%%=============================================================%%
%% GivenName	-> \fnm{Joergen W.}
%% Particle	-> \spfx{van der} -> surname prefix
%% FamilyName	-> \sur{Ploeg}
%% Suffix	-> \sfx{IV}
%% \author*[1,2]{\fnm{Joergen W.} \spfx{van der} \sur{Ploeg} 
%%  \sfx{IV}}\email{iauthor@gmail.com}
%%=============================================================%%

\author{\fnm{Makoto} \sur{Takeuchi}}\email{takeuchi\_makoto@cyberagent.co.jp}

%\author[2,3]{\fnm{Second} \sur{Author}}\email{iiauthor@gmail.com}
%\equalcont{These authors contributed equally to this work.}

%\author[1,2]{\fnm{Third} \sur{Author}}\email{iiiauthor@gmail.com}
%\equalcont{These authors contributed equally to this work.}

\affil{\orgname{CyberAgent, Inc.}, \orgaddress{\country{Japan}}}

%\affil[2]{\orgdiv{Department}, \orgname{Organization}, \orgaddress{\street{Street}, \city{City}, \postcode{10587}, \state{State}, \country{Country}}}

%\affil[3]{\orgdiv{Department}, \orgname{Organization}, \orgaddress{\street{Street}, \city{City}, \postcode{610101}, \state{State}, \country{Country}}}

%%==================================%%
%% Sample for unstructured abstract %%
%%==================================%%

\abstract{Music genre classification shapes how listeners discover music, how platforms design recommendations, and how sociologists study cultural taste. Yet existing genre labels are inconsistent in granularity: they exaggerate boundaries between overlapping categories and hide sociologically important heterogeneity within broad labels. Cultural sociologists have long theorized that genres vary along two independent dimensions, boundary strength and internal differentiation, but existing empirical work has relied on fixed label sets, leaving these dimensions without quantitative operationalization from actual consumption behavior data. Here we propose a two-dimensional framework that extracts music communities bottom-up from playlist co-occurrence networks and characterizes each along two axes: external closure $B(C)$, measuring boundary strength relative to a random null, and internal differentiation $D(C)$, measuring organized internal subdivision. We validate the framework on two independent datasets spanning different platforms, cultural contexts, and time periods, confirming that $B(C)$ and $D(C)$ are statistically independent and that each captures a distinct structural property. The framework reveals genre structures invisible to fixed labels: single labels splitting into communities with different boundary strengths, multiple labels merging into tightly bounded communities, and consumption spheres that no existing label describes. Comparison with prior theoretical predictions is broadly consistent, with the notable exception that Hip-Hop exhibits rich internal differentiation across both datasets, challenging its prevailing single-centered characterization. By providing a label-independent coordinate system grounded in listener behavior, this framework opens a path toward tracking how genre boundaries and internal structures evolve over time, a question that static label systems cannot address.
}


%%\pacs[JEL Classification]{D8, H51}

%%\pacs[MSC Classification]{35A01, 65L10, 65L12, 65L20, 65L70}

\maketitle

% ===================================================================
% 1. INTRODUCTION
% ===================================================================
\section{Introduction}

The classification of music genres has played a central role in the music industry, recommender systems, and cultural sociology. However, existing genre labels (Rock, Pop, Hip-Hop, Classical, etc.) are inconsistent in granularity: overly broad labels conceal internal heterogeneity, while overly narrow labels sever continuities in consumption behavior. Lizardo~\cite{Lizardo2024} criticized the macro-genre labels used in the sociology of taste as ``concealing as much as they reveal,'' identifying two problems. First, the \emph{boundary overlap problem}: although actual genres overlap and have ambiguous boundaries, surveys treat them as clear-cut divisions. Second, the \emph{internal heterogeneity problem}: sociologically important distinctions are hidden within broad labels.

Silver et al.~\cite{Silver2016} analyzed the co-occurrence network of self-reported genres by musicians on MySpace and presented a framework for organizing the structure of the music world along two axes: \emph{boundary strength} and \emph{internal differentiation}. Rock was characterized as high-boundary, high-differentiation (multi-centered); Hip-Hop as high-boundary, low-differentiation (single-centered); and Niche genres as low-boundary, high-differentiation (uncentered). This two-axis framework is theoretically significant in that it captures genres not as a single continuum but as combinations of two independent structural properties. However, Silver et al.'s analysis took 122 fixed labels as its starting point and thus did not escape label dependence. Moreover, the quantitative indicators for boundary strength and internal differentiation remained undeveloped, limited to qualitative description.

This study proposes a framework that constructs artist networks from playlist co-occurrences in a bottom-up manner, decomposes them into communities, treats each community $C$ as a music genre, and quantitatively describes its structure along two axes: \emph{external closure $B(C)$} and \emph{internal differentiation $D(C)$}. As a methodological stance, this study does not aim to reproduce existing genre labels as ``ground truth.'' Community extraction results are positioned as a ``device for discovering new clusters of music,'' the two indicators as a ``coordinate system for comparing and describing those clusters,'' and existing labels as ``annotations overlaid post hoc.'' This approach gives rise to three research questions:

\begin{itemize}
    \item \textbf{RQ1 (Framework validity)}: Can the community structure of playlist co-occurrence networks be meaningfully characterized along two independent dimensions, external closure $B(C)$ and internal differentiation $D(C)$?
    \item \textbf{RQ2 (Theoretical correspondence)}: To what extent does the resulting two-dimensional landscape align with, or deviate from, the genre typology qualitatively described by Silver et al.~\cite{Silver2016}?
    \item \textbf{RQ3 (Discovery beyond labels)}: What genre structures, invisible to existing label systems, does the bottom-up approach reveal?
\end{itemize}

\noindent RQ1 is addressed through three statistical hypotheses (H1--H3) in Section~5.2. RQ2 is examined through comparison with Silver et al.'s theoretical predictions in Section~5.3. RQ3 is addressed through case studies in Section~5.4.

The contributions of this study are summarized in three points. (1)~\emph{Measurement contribution}: We quantify the boundary strength and internal differentiation that Silver et al.~\cite{Silver2016} discussed qualitatively, operationalizing them as the Conductance Deficit relative to a Configuration Model ($B(C)$) and recursive subcommunity decomposition ($D(C)$), and propose a coordinate system that positions communities on a two-dimensional plane. (2)~\emph{Validation contribution}: We confirm the validity of the proposed indicators through three statistical hypotheses: $B(C)$ and $D(C)$ are independent after controlling for size (H1), $D(C)$ correlates with entropy reduction of existing genre labels (H2), and $B(C)$ correlates with the concentration of external connection patterns (H3). We further examine the correspondence between the observed $B \times D$ landscape and Silver et al.~\cite{Silver2016}'s theoretical predictions. (3)~\emph{Discovery contribution}: Through the bottom-up approach, we reveal structures invisible to existing labels: cases where a single label splits into multiple communities with different $B(C)$ values, cases where multiple labels merge into a single high-$B(C)$ community, and internal differentiation of Hip-Hop that is consistent across cultural contexts.

This study also addresses the ``incompleteness and inconsistency of genre metadata'' that the MIR (Music Information Retrieval) field has long recognized \cite{SchedlGomez2014, Bogdanov2019} by providing a structural description framework that does not depend on fixed labels. The approach of extracting and quantifying genre structure bottom-up from consumption behavior data (playlist co-occurrences) is also applicable to complementing and enriching genre metadata in recommender systems and playlist curation.

% ===================================================================
% 2. RELATED WORK
% ===================================================================
\section{Related Work}

\subsection{Classification theory of music genres}

DiMaggio~\cite{DiMaggio1987} argued that classification systems in the arts should be understood along multiple dimensions, including \emph{differentiation} (the degree of subdivision) and \emph{ritual strength} (the strength of boundaries). Silver et al.~\cite{Silver2016} applied this framework to music genres, organizing the music world along two axes of boundary strength and internal differentiation. Importantly, Silver et al.\ explicitly stated that the criteria for subdivision should be localized to each subcluster rather than applied uniformly across the whole, theoretically supporting our recursive decomposition approach.

Lizardo~\cite{Lizardo2024} criticized macro-genre labels for concealing \emph{boundary overlap} and \emph{internal heterogeneity}, and proposed extracting focused microgenres through link clustering. Lizardo's argument, at least conceptually, treats the exaggeration of boundaries and the concealment of internal heterogeneity as separate problems, aligning with our separation of $B(C)$ and $D(C)$.

\subsection{Category spanning and genre boundaries}

Kov\'{a}cs and Hannan~\cite{KovacsHannan2015} argued for the need to distinguish and separately measure category \emph{contrast} and \emph{niche width}. Shi et al.~\cite{Shi2018} demonstrated that genre crossing incurs stronger penalties when it spans nonporous, high-contrast boundaries. These findings provide the theoretical basis for our expectation (H3) that external adjacencies of high-$B(C)$ communities are biased toward ``near exterior'' rather than ``distant exterior.''

\subsection{Genre metadata challenges in MIR}

In the field of Music Information Retrieval (MIR), the incompleteness of genre metadata has long been recognized as a challenge. The comprehensive survey by Schedl et al.~\cite{SchedlGomez2014} noted that genre labels differ in granularity and criteria across information sources, with no consistent classification system in existence. Bogdanov et al.~\cite{Bogdanov2019} empirically demonstrated at scale that four sources (AllMusic, Discogs, Last.fm, and Tagtraum) assign entirely different genre hierarchies to the same tracks. Schreiber~\cite{Schreiber2015} reported that 30--40\% of genre annotations are missing even in the Million Song Dataset. Lamere~\cite{Lamere2008} showed that Last.fm folksonomy tags follow a long-tail distribution, making them difficult to apply to minor genres.

This problem directly affects the quality of music recommendation. Celma~\cite{Celma2010} demonstrated that popularity bias in recommender systems impedes the reach of niche music on the long tail, arguing that understanding genre structure is important for addressing this issue. Jiang et al.~\cite{JiangSpotify2024} used large-scale Spotify data to empirically show that consumption patterns differ structurally by genre, suggesting that quantitative understanding of genre structure directly contributes to improving recommendation and curation.

Attempts at bottom-up discovery of genre structure also exist. Corr\^{e}a et al.~\cite{CorreaDe2011} constructed networks from acoustic features and detected genre communities without predefined labels. Jiang and Huynh~\cite{JiangHuynh2022} proposed an approach for discovering genre structure bottom-up from user review behavior, though their scope was limited to metal music subgenres, precluding cross-genre comparison of boundary strength. However, none of these studies systematically quantified the \emph{structural properties} of discovered communities, neither the strength of boundaries nor the degree of internal differentiation, nor did they verify the correspondence between structural metrics and label-based heterogeneity. This study extends these prior works by evaluating bottom-up extracted communities along the two-dimensional $B(C) \times D(C)$ framework.

\subsection{Community structure in music networks}

Several studies have analyzed community structure in music networks and mapped communities to genres or musical styles. Park and Park~\cite{ParkPark2025} showed that communities in sampling networks correspond to musical styles, and Park et al.~\cite{ParkBaeSchichPark2015} showed that communities in co-recording networks correspond to historical periods of music. Regarding playlist data specifically, McFee and Lanckriet~\cite{McFee2011, McFee2012} empirically demonstrated that track co-occurrence within playlists reflects musical relationships, and noted that user-generated playlists possess different characteristics from those produced by recommendation services or commercial radio~\cite{McFee2011}; by extension, playlists curated by users are less likely to be confounded by platform recommendation algorithms, making them useful data for extracting music genre information bottom-up from user behavior. Against this background, this study also employs playlist data generated by music listeners. None of these studies systematically quantified the structural properties of the communities they discovered, neither the strength of boundaries nor the degree of internal differentiation, nor did they compare these properties across communities within the same network.

\subsection{The scale problem in network science}

Community detection suffers from a resolution limit \cite{Fortunato2007}, and multiple topological scales can coexist within a network \cite{Arenas2008}. Leskovec et al.~\cite{Leskovec2008} showed that in large-scale networks, community conductance depends on scale, and communities of different boundary qualities coexist at any given resolution snapshot. In this study, we understand the granularity mismatch between existing labels and communities as a natural consequence of this scale problem.

% ===================================================================
% 3. CONCEPTUAL FRAMEWORK
% ===================================================================
\section{Conceptual Framework}

\subsection{Approach overview and scope}

This study adopts a network-topology-based approach: communities extracted from a playlist co-occurrence network are characterized along two dimensions (external closure $B(C)$ and internal differentiation $D(C)$), and the resulting two-dimensional landscape is examined for systematic patterns. Several methodological commitments follow from this design.

First, the communities we analyze are bottom-up structures defined by network topology, not predefined genre categories. Existing genre labels are used solely as \emph{post-hoc annotations} for interpreting the discovered structure, not as ground truth to be recovered. We therefore do not evaluate partitions by their agreement with any particular label set, nor do we claim that the extracted communities represent the only valid segmentation of the music space.

\subsection{Unit of analysis: First-level communities}

The basic unit of our two-dimensional mapping is the \emph{first-level community} obtained by partitioning the full network once. Subcommunities are used primarily to compute $D(C)$, and we do not place parent and child communities side by side on the same two-dimensional map.

\subsection{External closure $B(C)$}

$B(C)$ measures the extent to which a first-level community $C$ is contained within itself rather than leaking to the exterior of the overall network, corresponding to the boundary strength concept of Silver et al.~\cite{Silver2016}.

We operationalize this as the \emph{Conductance Deficit} (CD) relative to a Configuration Model (a random graph preserving the original degree sequence):

\begin{equation}
\text{CD}(C) = E[\phi_{\text{random}}] - \phi_{\text{observed}}
\end{equation}

where $\phi$ is conductance $\phi(S) = \text{cut}(S, \bar{S}) / \min(\text{vol}(S), \text{vol}(\bar{S}))$, and $E[\phi_{\text{random}}]$ is the average over 100 Configuration Model simulations. A higher CD indicates a stronger boundary relative to random expectation.

The absolute value of CD depends on network density: in denser networks, the structural lower bound of observed conductance is higher, resulting in lower absolute CD values overall. Therefore, the high/low classification of $B(C)$ is based on relative rank order within each dataset, and cross-dataset comparison of absolute values is not performed.

\subsection{Internal differentiation $D(C)$}

$D(C)$ measures the extent to which a first-level community $C$ subdivides into organized subcommunities when recursively partitioned, corresponding to the internal differentiation concept of Silver et al.~\cite{Silver2016}.

Following Silver et al.'s methodology, we apply the Leiden algorithm (resolution $\gamma = 1.0$) within each community and evaluate the significance of each partition using the Wilcoxon signed-rank test (internal degree vs.\ external degree for each node, $p < 0.01$). $D(C)$ is defined as the number of significant subcommunities.

Larger communities are more likely to contain substructures, and $D(C)$ is expected to have a positive correlation with community size. The problem here is that the statistical power of the Wilcoxon test also depends on size. We address this in H1 ($B$--$D$ independence test) by treating size as a control variable.

\subsection{Interpretation of the two-dimensional mapping}

The four quadrants corresponding to Silver et al.~\cite{Silver2016}'s theoretical framework:

\begin{table}[h]
\centering
\caption{Four quadrants of the $B(C) \times D(C)$ framework, corresponding to Silver et al.~\cite{Silver2016}'s genre typology.}
\label{tab:quadrants}
\begin{tabular}{lll}
\toprule
 & High $D(C)$ & Low $D(C)$ \\
\midrule
\textbf{High $B(C)$} & Multi-centered (strong boundary, & Single-centered / Focused (strong \\
 & many substructures) & boundary, homogeneous interior) \\
\textbf{Low $B(C)$} & Uncentered (weak boundary, & Free interchangeability (weak \\
 & many substructures) & boundary, homogeneous interior) \\
\bottomrule
\end{tabular}\end{table}

This study is positioned not as a strict replication of Silver et al., but as an empirical investigation that transplants the two-axis theoretical framework to listener-behavior-based networks and tests it against bottom-up extracted communities.

% ===================================================================
% 4. DATA AND METHODS
% ===================================================================
\section{Data and Methods}

\subsection{Datasets}

Two datasets are used to test the generalizability of the proposed framework across platforms, cultural contexts, and time periods.

\textbf{AWA (primary dataset):}

AWA~\cite{AWA} is a Japanese music streaming service. The AWA dataset consists of user-curated playlists randomly sampled at a fixed rate from playlists with playback records during the observation period on the platform. The target playlists are limited to those with public settings accessible to other users, in order to exclude personal music-logging playlists.

\begin{table}[h]
\centering
\caption{AWA dataset summary.}
\begin{tabular}{ll}
\toprule
Item & Value \\
\midrule
Platform & Japanese music streaming service (AWA) \\
Period & 2023 (one year) \\
Playlist length & Fixed at 8 tracks \\
Nodes (total) & 21,665 artists \\
Edge definition & Co-occurrence count within the same playlist \\
Co-occurrence threshold & $\geq 50$ \\
LCC (after thresholding) & 2,779 nodes, 24,298 edges \\
Network density & 0.006 \\
Genre labels & 13 categories (service-defined), artist-level, coverage 75.3\% \\
\bottomrule
\end{tabular}\end{table}

\textbf{AotM-2011:}

The AotM-2011 dataset is a complete crawl of all playlists posted on the Art of the Mix website from its inception through mid-2011, compiled by McFee and Lanckriet~\cite{McFee2012}. Song identifiers were resolved against the Million Song Dataset (MSD); playlists were then segmented into contiguous runs of matched tracks.

\begin{table}[h]
    \centering
    \caption{AotM-2011 dataset summary.}
    \begin{tabular}{ll}
    \toprule
    Item & Value \\
    \midrule
    Platform & Art of the Mix (English-language playlist-sharing site) \\
    Period & 1998--2011 \\
    Number of playlists & 101,343 \cite{McFee2012} \\
    Playlist length (original median) & 19 tracks \\
    Playlist-length normalization & First 8 tracks \\
    Co-occurrence threshold & $\geq 4$ \\
    LCC (after normalization and thresholding) & 4,671 nodes, 75,948 edges \\
    Network density & 0.007 \\
    Genre labels & 43 categories (user self-reported, playlist-level only) \\
    \bottomrule
    \end{tabular}\end{table}

\textbf{Playlist-length normalization.} AWA playlists are fixed at 8 tracks due to a platform feature constraint (the public playlist function limits playlist length). AotM playlists are substantially longer (median 19 tracks), generating 5.9 times more co-occurrence pairs per playlist ($\binom{17}{2} = 136$ vs.\ $\binom{8}{2} = 28$). To prevent the resulting network from becoming excessively dense and thereby suppressing community structure, we truncate each AotM playlist to its first 8 tracks. AotM playlists are user-curated mixtapes with intentional track ordering, so the first-8 selection is deterministic and reproducible.

\textbf{Co-occurrence threshold calibration.} Even after playlist-length normalization, the artist population sizes differ substantially (AWA: 21,665; AotM after truncation: 103,801), causing the expected co-occurrence count per pair under a random null to differ by a factor of 56 (AWA: $\lambda = 0.029$; AotM: $\lambda = 0.0005$). We therefore calibrate the co-occurrence threshold so that the resulting network density is comparable across datasets (AWA: $\geq 50$, density 0.006; AotM: $\geq 4$, density 0.007). Details of this calibration procedure and sensitivity analyses with alternative thresholds are reported in Appendix~\ref{app:preprocessing}.

The two datasets differ in platform, cultural context, and time period. After the two-step normalization described above, network density is comparable (AWA: 0.006, AotM: 0.007). The two datasets also differ in the nature of their genre metadata. AWA provides artist-level genre labels (13 service-defined categories), while AotM provides only playlist-level categories (43 user self-reported categories such as ``Rock/Pop,'' ``Hip Hop,'' and ``Indie Rock''). For AotM, community-level genre profiles are estimated by aggregating the playlist categories of the artists within each community. Since AotM lacks artist-level labels, analyses requiring per-artist labels (e.g., H2) are conducted on AWA only.

\subsection{Network construction and community extraction}

For each dataset, we analyze the largest connected component (LCC) of the artist co-occurrence network after applying the co-occurrence threshold.

Community extraction uses the Leiden algorithm \cite{Traag2019} with resolution parameter $\gamma = 1.0$. We adopt the Leiden algorithm over the widely used Louvain method for two reasons. First, Leiden guarantees that every detected community is internally connected, whereas Louvain can produce badly connected communities \cite{Traag2019}; this guarantee is a prerequisite for our recursive decomposition of communities in the $D(C)$ computation. Second, Leiden's faster convergence makes it well suited to the consensus clustering pipeline, which requires a large number of repeated runs. A $\gamma$ sweep (0.3--5.0) confirmed that major communities emerge stably in the range $\gamma = 0.5$--$1.5$.

To address seed dependence of the Leiden algorithm, we applied the iterative consensus clustering method of Lancichinetti and Fortunato~\cite{Lancichinetti2012} to the AWA data. A co-assignment matrix was constructed from 100 Leiden runs, converted to a consensus network at threshold $\tau = 0.5$, and the Leiden algorithm was reapplied iteratively until convergence. Full convergence was achieved in one iteration, yielding a stable partition (NMI $= 1.000$) reproducible with any seed.

As a result, 65 communities were extracted for AWA (29 with size $\geq 5$) and 17 communities for AotM (13 with size $\geq 5$).

The AWA network after thresholding (4,374 nodes) consists of 685 connected components, with the LCC comprising 2,779 nodes (63.5\%). The remaining 1,595 nodes (36.5\%) are distributed across 684 non-LCC components. Most are micro-components of size 2--3, but 19 components of size $\geq 5$ exist. Non-LCC components have zero connections to the LCC, representing the theoretical limit of $B(C)$, i.e., completely isolated consumption spheres. These structures are analyzed in Section~5.4.

\subsection{Computation of $B(C)$}

For each first-level community, we computed the observed conductance $\phi_{\text{observed}}$ and calculated the Conductance Deficit (CD) as the difference from the expected conductance $E[\phi_{\text{random}}]$ obtained from 100 Configuration Model simulations.

\subsection{Computation of $D(C)$}

For the internal subnetwork of each first-level community, subcommunities were detected using the Leiden algorithm ($\gamma = 1.0$), and the significance of each subcommunity was evaluated by the Wilcoxon signed-rank test ($p < 0.01$). $D(C)$ is defined as the number of significant subcommunities. The minimum subcommunity size was set to 5. This method follows the recursive decomposition approach of Silver et al.~\cite{Silver2016}, with the algorithm replaced by the Leiden method.

\subsection{Hypothesis testing and research question design}

To address the three research questions posed in Section~1, we employ the following design.

\textbf{RQ1 (Framework validity)} is tested through three statistical hypotheses:

\begin{itemize}
    \item \textbf{H1 ($B \perp D$ independence)}: $B(C)$ and $D(C)$ are independent after controlling for community size. Tested via Spearman partial correlation.
    \item \textbf{H2 (Entropy reduction)}: Communities with higher $D(C)$ show greater reduction in Shannon entropy of existing genre labels ($\Delta H$) after subcommunity decomposition. Conducted on AWA only (AotM lacks artist-level labels).
    \item \textbf{H3 (Concentration of external adjacency)}: Communities with higher $B(C)$ have external adjacent artists whose connections are concentrated in fewer communities (lower external community entropy). Conducted on both datasets.
\end{itemize}

\textbf{RQ2 (Theoretical correspondence)} is addressed by comparing the $B \times D$ positions of major genre communities against Silver et al.~\cite{Silver2016}'s predictions (Rock = multi-centered, Hip-Hop = single-centered, Niche = uncentered). This comparison is not framed as a statistical hypothesis test but as a systematic qualitative assessment.

\textbf{RQ3 (Discovery beyond labels)} is addressed through case studies that examine structures revealed by the bottom-up approach that are invisible to existing label systems.

\textbf{Multiple comparison correction.} Six primary statistical tests are conducted across H1--H3 (two partial correlations for H1, one Spearman correlation and one dichotomized test for H2, and two Spearman correlations for H3). We apply Benjamini--Hochberg (BH) false discovery rate correction to account for multiple testing.

% ===================================================================
% 5. RESULTS
% ===================================================================
\section{Results}

\subsection{Overview and two-dimensional mapping}

For AWA, consensus clustering on the LCC (2,779 nodes) yielded 65 communities (modularity $Q = 0.715$). The size distribution is heavily right-skewed, with the top 7 communities accounting for 83.3\% of all nodes. The analysis sample (size $\geq 5$) comprises 29 communities.

\begin{figure}
\centering
\includegraphics[width=\textwidth]{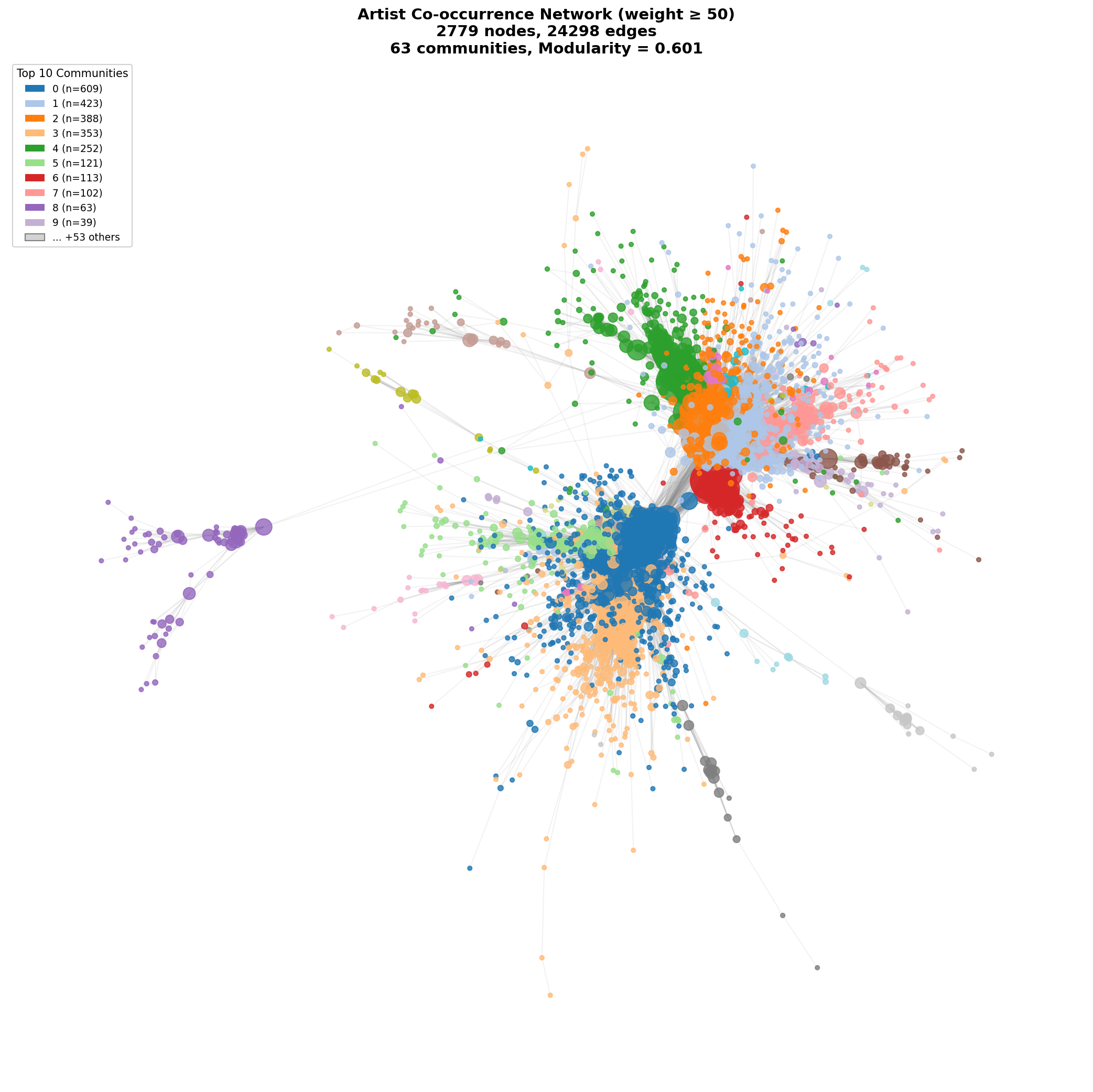}
\caption{Artist co-occurrence network for AWA (LCC, co-occurrence threshold $\geq 50$, 2,779 nodes, 24,298 edges). Each node represents an artist; edges connect artists who co-occur in $\geq 50$ playlists. Node colors indicate communities detected by consensus clustering with the Leiden algorithm (top 10 communities shown in legend; 53 smaller communities shown in grey). Node size is proportional to degree. Layout: ForceAtlas2. Modularity $Q = 0.601$.}
\label{fig:network}
\end{figure}

Outside the LCC, 684 connected components exist (totaling 1,595 nodes, 36.5\% of the network). Most are micro-components of size 2--3 (573 + 75 = 648 components), but 19 components of size $\geq 5$ are included. The genre composition of non-LCC components differs substantially from the LCC, with higher proportions of soundtrack (14.2\%) and dance-electronic (15.0\%) compared to within the LCC. The existence of multiple pure-soundtrack isolated components suggests that certain genre consumption spheres are structurally separated from the mainstream (discussed in detail in Section~5.4).

For AotM (after playlist-length normalization and threshold calibration), 17 communities were extracted ($Q = 0.308$). The analysis sample (size $\geq 5$) comprises 13 communities.

\begin{figure}
\centering
\includegraphics[width=\textwidth]{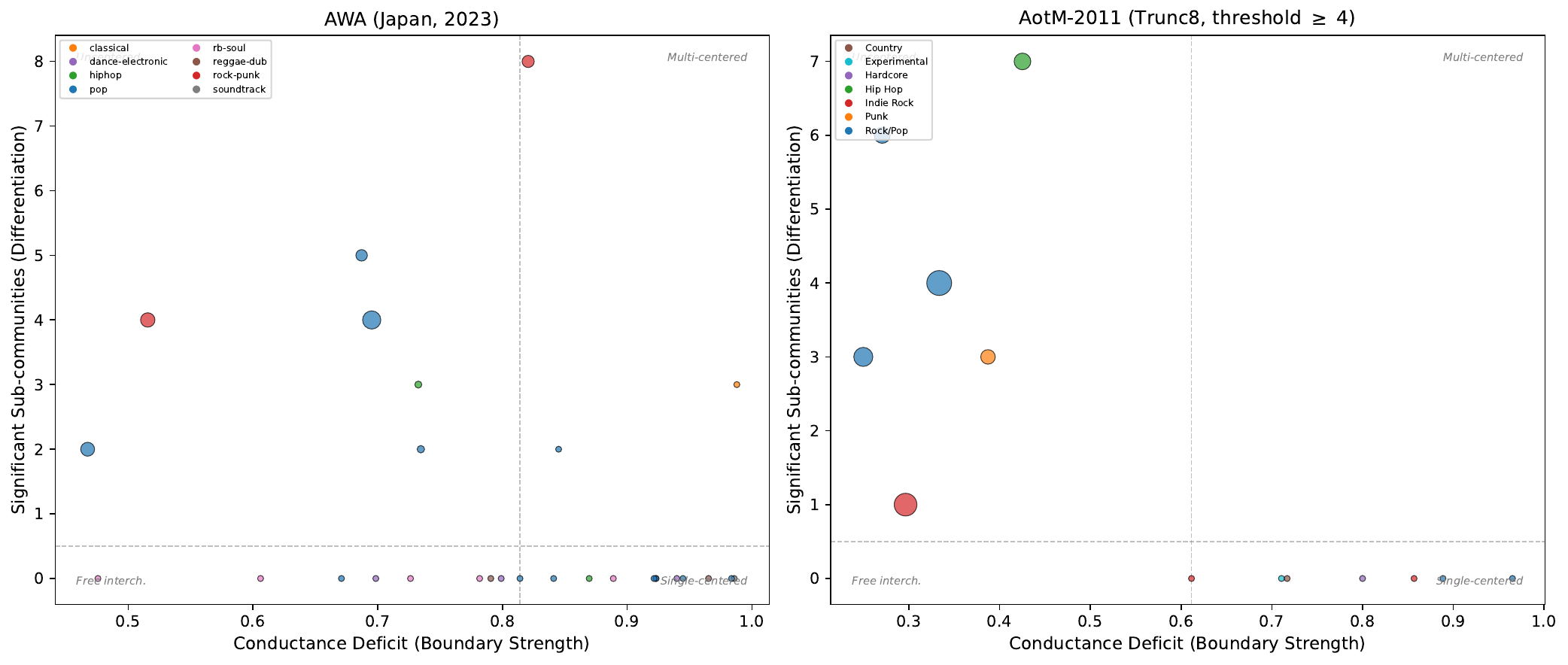}
\caption{Two-dimensional $B(C) \times D(C)$ mapping for AWA (left) and AotM (right). Horizontal axis: external closure $B(C)$, operationalized as Conductance Deficit (higher = stronger boundary). Vertical axis: internal differentiation $D(C)$, measured as the number of significant subcommunities (higher = more internally differentiated). Node size is proportional to community size. Node color indicates the primary genre label of each community (see legend). Dashed lines indicate the median $B(C)$ threshold used for high/low classification within each dataset. The four quadrants correspond to Silver et al.~\cite{Silver2016}'s typology: Multi-centered (high B, high D), Single-centered (high B, low D), Uncentered (low B, high D), and Free interchangeability (low B, low D). }
\label{fig:2d_mapping}
\end{figure}

Figure~\ref{fig:2d_mapping} shows the $B(C) \times D(C)$ landscape for both datasets. In AWA, communities are distributed across all four quadrants of the Silver et al.~\cite{Silver2016} typology (Table~\ref{tab:quadrants}), confirming that the two dimensions produce meaningful variation rather than collapsing onto a single axis. Large communities tend to have high $D(C)$ and low-to-moderate $B(C)$, while small communities cluster in the high-$B(C)$, low-$D(C)$ region. In AotM, the same size-dependent pattern is observed: all 6 communities with $D(C) > 0$ are among the largest (size $\geq 431$), while the 7 smaller communities (size $\leq 23$) all have $D(C) = 0$. $B(C)$ ranges from 0.25 to 0.97 in AotM, with small communities exhibiting higher $B(C)$ and large communities lower $B(C)$, paralleling AWA. Note that because the absolute value of CD depends on network density, $B(C)$ values are not directly comparable across datasets; the high/low classification is based on relative rank order within each dataset.

\textbf{Distribution of $D(C)$:} The frequency distributions of $D(C)$ are as follows:

\begin{table}
\centering
\caption{Frequency distribution of $D(C)$ across datasets.}
\begin{tabular}{lll}
\toprule
$D(C)$ & AWA ($n=29$) & AotM ($n=13$) \\
\midrule
0 & 20 (69.0\%) & 7 (53.8\%) \\
1 & 0 & 1 (7.7\%) \\
2 & 3 (10.3\%) & 0 \\
3 & 2 (6.9\%) & 2 (15.4\%) \\
4 & 2 (6.9\%) & 1 (7.7\%) \\
5 & 1 (3.4\%) & 0 \\
6 & 0 & 1 (7.7\%) \\
7 & 0 & 1 (7.7\%) \\
8 & 1 (3.4\%) & 0 \\
\bottomrule
\end{tabular}\end{table}

$D(C)$ is a zero-inflated discrete variable in both datasets: 69\% of AWA communities and 54\% of AotM communities have $D(C) = 0$. Communities with $D(C) > 0$ are all among the larger communities (AWA: size $\geq 63$; AotM: size $\geq 431$), consistent with the size dependence noted in Section~3.4. Given this zero-inflated distribution, we supplement Spearman correlations involving $D(C)$ with dichotomized analyses ($D(C) > 0$ vs.\ $D(C) = 0$) in the hypothesis tests below.

\subsection{RQ1: Validity of the two-dimensional framework}

We test three hypotheses to establish that $B(C)$ and $D(C)$ constitute a meaningful two-dimensional coordinate system. All results that are significant at $\alpha = 0.05$ before multiple comparison correction remain significant after Benjamini--Hochberg (BH) correction; the most conservative Bonferroni correction also preserves all significant results.

\subsubsection{H1: Independence of $B(C)$ and $D(C)$}

As a prerequisite for the two-dimensional framework, we confirm that $B(C)$ and $D(C)$ capture distinct structural properties.

\begin{table}
\centering
\begin{tabular}{lll}
\toprule
Test item & AWA ($n=29$) & AotM ($n=13$) \\
\midrule
$\rho(B, D)$ zero-order & $-0.304$ & $-0.778${**} \\
$\rho(B, \log\text{size})$ & $-0.329$ & $-0.821${***} \\
$\rho(D, \log\text{size})$ & $0.782${***} & $0.795${***} \\
$\rho(B, D \mid \log\text{size})$ partial & $-0.080$ & $-0.361$ \\
\bottomrule
\end{tabular}
\caption{H1: Independence test results. Significance levels: {***} $p < 0.001$, {**} $p < 0.01$, {*} $p < 0.05$.}
\label{tab:h1}
\end{table}

$B(C)$ and $D(C)$ are independent after controlling for community size in both datasets (Table~\ref{tab:h1}). In AWA, both indicators correlate with size ($D(C)$ positively, $B(C)$ negatively), producing an apparent negative $B$--$D$ correlation. After partialing out size, the correlation drops to $-0.080$ ($p = 0.685$). In AotM, size confounding is even stronger, yet the partial correlation is again non-significant ($-0.361$, $p = 0.248$). This result demonstrates that the two-dimensional independence theoretically assumed by DiMaggio~\cite{DiMaggio1987} and indirectly supported by Silver et al.~\cite{Silver2016} holds for communities on playlist co-occurrence networks across both datasets.

\subsubsection{H2: Entropy reduction through internal differentiation}

If $D(C)$ is a meaningful indicator, recursive partitioning of a community's interior should sort the mixed existing genre labels into more ordered groupings. For AWA's 29 communities (size $\geq 5$, label coverage $\geq 50\%$), we computed the difference ($\Delta H$) between Shannon entropy before partitioning and the size-weighted average entropy after partitioning.

\begin{figure}
\centering
\includegraphics[width=\textwidth]{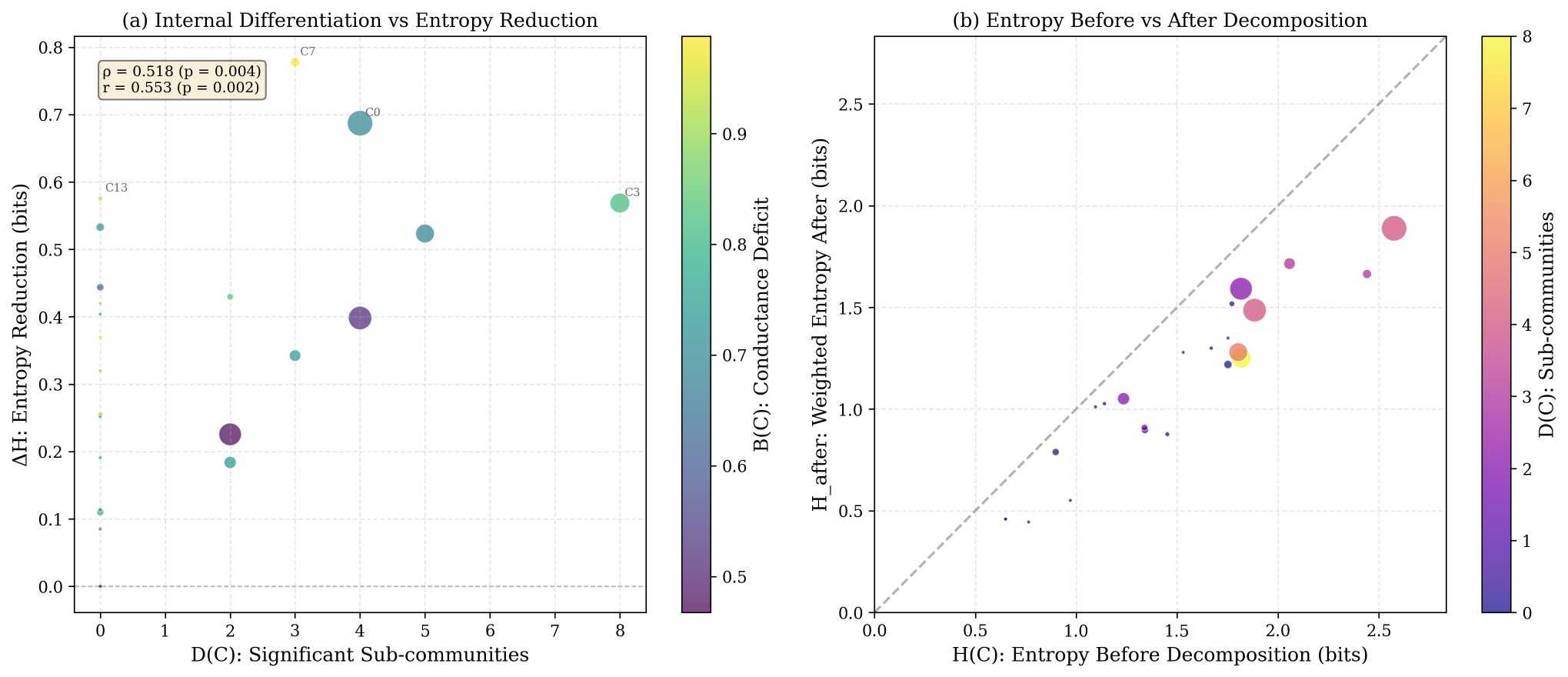}
\caption{Relationship between internal differentiation $D(C)$ and entropy reduction (AWA, $n = 29$). \textbf{(a)} Scatter plot of $D(C)$ (x-axis: number of significant subcommunities) vs.\ $\Delta H$ (y-axis: Shannon entropy reduction in bits after subcommunity decomposition). Point size is proportional to community size; color indicates $B(C)$ (Conductance Deficit). Spearman $\rho = 0.518$**. \textbf{(b)} Entropy before decomposition $H(C)$ (x-axis) vs.\ size-weighted entropy after decomposition (y-axis). Points below the diagonal indicate entropy reduction; color indicates $D(C)$. Communities with higher $D(C)$ show greater displacement below the diagonal.}
\label{fig:entropy}
\end{figure}

\begin{table}
\centering
\begin{tabular}{ll}
\toprule
Test item & Result \\
\midrule
$\rho(D(C), \Delta H)$ & $0.518${**} \\
$r(D(C), \Delta H)$ & $0.553${**} \\
$\rho(B(C), \Delta H)$ & $-0.050$ \\
$\Delta H$: $D(C)>0$ vs $D(C)=0$ & 0.460 vs 0.213 bits{**} \\
Mean entropy & $H(C) = 1.187 \rightarrow H_{\text{after}} = 0.897$ bits \\
\bottomrule
\end{tabular}
\caption{H2: Entropy reduction test results (AWA, $n = 29$). {***} $p < 0.001$, {**} $p < 0.01$, {*} $p < 0.05$.}
\label{tab:h2}
\end{table}

$D(C)$ correlates with entropy reduction ($\rho = 0.518$**), while $B(C)$ does not (Table~\ref{tab:h2}), indicating that entropy reduction is an effect specific to $D(C)$. A dichotomized analysis confirms this pattern: communities with $D(C) > 0$ show significantly greater $\Delta H$ than those with $D(C) = 0$ (0.460 vs.\ 0.213 bits**). $D(C)$ thus captures organized internal structure, not mere variance. H2 is conducted on AWA only, as AotM lacks artist-level genre labels.

\subsubsection{H3: External closure and external connection patterns}

If $B(C)$ captures not merely the rate of external leakage but the quality of boundary closure, communities with higher $B(C)$ should have external connections concentrated in a small number of neighboring communities rather than dispersed broadly. We tested this by computing the Shannon entropy of external neighbors' destination community distribution for each community.

\begin{figure}
\centering
\includegraphics[width=\textwidth]{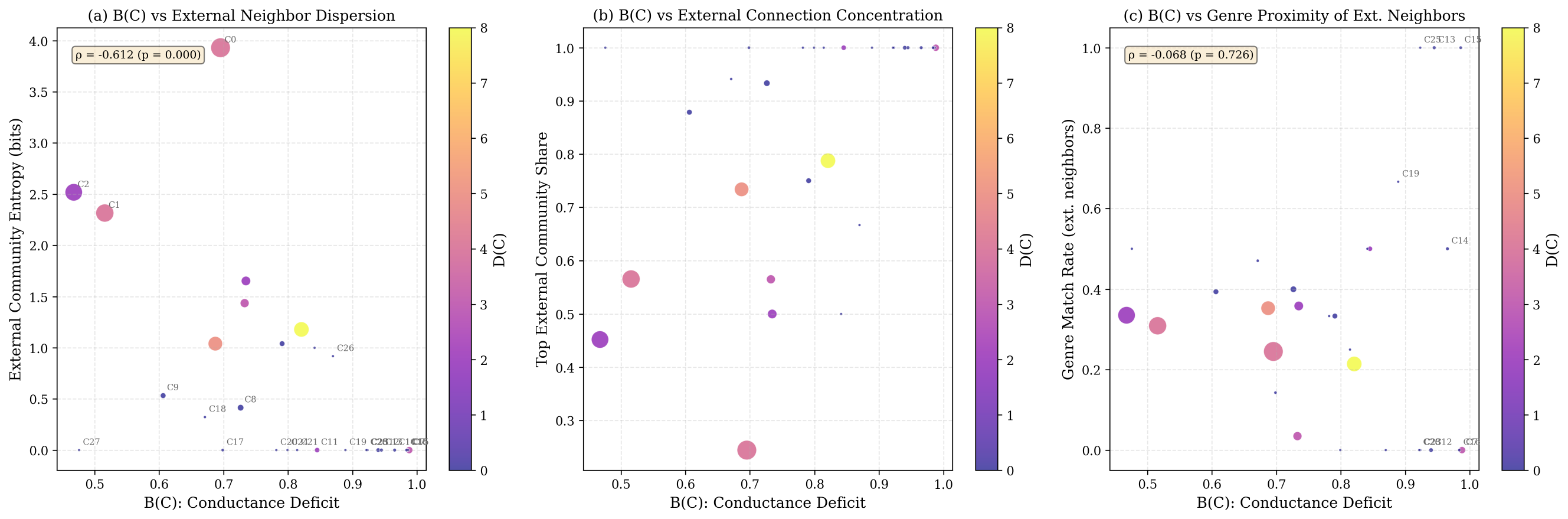}
\caption{Relationship between external closure $B(C)$ and external connection patterns (AWA, $n = 29$). \textbf{(a)} $B(C)$ (x-axis: Conductance Deficit) vs.\ external community entropy (y-axis: Shannon entropy of the distribution of external neighbors across communities; lower = more concentrated). $\rho = -0.612$***. \textbf{(b)} $B(C)$ vs.\ top external community share (y-axis: proportion of external neighbors connected to the single most frequent external community; higher = more concentrated). \textbf{(c)} $B(C)$ vs.\ genre match rate of external neighbors (y-axis: proportion of external neighbors sharing the same primary genre label; higher = genre-proximate). $\rho = -0.968$***. Point size is proportional to community size; color indicates $D(C)$.}
\label{fig:external}
\end{figure}

\begin{table}
\centering
\begin{tabular}{lll}
\toprule
Test item & AWA ($n=29$) & AotM ($n=13$) \\
\midrule
$\rho(B(C), \text{ext. comm. entropy})$ & $-0.612${***} & $-0.853${***} \\
$\rho(B(C), \text{ext. comm. HHI})$ & $0.602${***} & $0.875${***} \\
$\rho(B(C), n_{\text{ext. communities}})$ & $-0.638${***} & $-0.934${***} \\
Partial $\rho(B(C), \text{ext. entropy} \mid \text{size})$ & $-0.557${***} & $-0.560$ \\
$\rho(B(C), \text{ext. genre entropy})$ & $-0.729${***} & -- \\
\bottomrule
\end{tabular}
\caption{H3: External connection pattern test results. {***} $p < 0.001$, {**} $p < 0.01$, {*} $p < 0.05$. AotM ext.\ genre entropy is unavailable due to the absence of artist-level labels.}
\label{tab:h3}
\end{table}

Communities with higher $B(C)$ have external connections concentrated in fewer communities in both datasets (Table~\ref{tab:h3}). This effect is robust after controlling for size (AWA: partial $\rho = -0.557$***; AotM: partial $\rho = -0.560$, $p = 0.058$). The consistency across datasets strengthens the finding that high-$B(C)$ communities exhibit directional connection patterns limited to ``near exterior'' rather than arbitrary exterior, consistent with Silver et al.~\cite{Silver2016}'s finding that Rock and Hip-Hop have low bridge ratios to the world outside, and supporting Shi et al.~\cite{Shi2018}'s argument that crossing nonporous boundaries incurs larger penalties.

\subsection{RQ2: Comparison with Silver et al.'s (2016) theoretical predictions}

To assess how the observed two-dimensional landscape corresponds to prior theoretical work, we compared the $B \times D$ positions of major genre communities against Silver et al.'s predictions. This comparison is not a statistical hypothesis test but a systematic qualitative assessment. The high/low threshold for $B(C)$ was the median (0.814) within the AWA dataset, with $D(C) > 0$ classified as ``high D'' and $D(C) = 0$ as ``low D.'' Communities for which Silver et al.\ did not provide an explicit prediction are marked as ``Ref.'' (reference only). Table~\ref{tab:rq2_awa} shows the results for AWA, and Table~\ref{tab:rq2_aotm} for AotM.

\begin{table}
\centering
\scriptsize
\begin{tabular}{llrrlll}
\toprule
Genre & Size & $B(C)$ & $D(C)$ & Prediction & Observed & Agreement \\
\midrule
rock-punk (C3) & 302 & 0.821 & 8 & high B, high D & high B, high D & Match \\
rock-punk (C1) & 423 & 0.516 & 4 & high B, high D & low B, high D & D match \\
hiphop (C6) & 102 & 0.733 & 3 & high B, low D & low B, high D & Mismatch \\
pop (C0) & 679 & 0.695 & 4 & low B & low B, high D & B match \\
classical (C7) & 63 & 0.988 & 3 & -- & high B, high D & Ref. \\
reggae-dub (C14) & 16 & 0.965 & 0 & high B, low D & high B, low D & Match \\
dance-elec. (C12) & 24 & 0.940 & 0 & low B & high B, low D & Opposite \\
\bottomrule
\end{tabular}
\caption{AWA: Comparison with Silver et al.~\cite{Silver2016}'s predictions.}
\label{tab:rq2_awa}
\end{table}

\begin{table}
\centering
\scriptsize
\begin{tabular}{llrrlll}
\toprule
Genre & Size & $B(C)$ & $D(C)$ & Prediction & Observed & Agreement \\
\midrule
Rock/Pop (C0) & 1,261 & 0.334 & 4 & high B, high D & mid B, high D & D match \\
Hip Hop (C3) & 569 & 0.425 & 7 & high B, low D & high B, high D & Mismatch \\
Indie Rock (C1) & 1,049 & 0.297 & 1 & -- & low B, high D & Ref. \\
Punk (C5) & 431 & 0.387 & 3 & -- & mid B, high D & Ref. \\
Country (C6) & 23 & 0.717 & 0 & -- & high B, low D & Ref. \\
\bottomrule
\end{tabular}
\caption{AotM: Comparison with Silver et al.~\cite{Silver2016}'s predictions. Genre labels are derived from playlist-level categories (see Section~4.1).}
\label{tab:rq2_aotm}
\end{table}

The comparison reveals broad agreement with Silver et al.'s predictions alongside informative deviations. Rock communities show high $D(C)$ in both datasets, confirming that Rock's multi-centered structure holds across platforms and cultural contexts. Small-scale communities (Country, Hardcore) consistently exhibit high $B(C)$ and low $D(C)$. Hip-Hop, however, shows the highest $D(C)$ in both datasets (AWA: $D=3$; AotM: $D=7$), consistently deviating from Silver et al.'s single-centered prediction. This aligns with Lizardo~\cite{Lizardo2024}'s observation that hidden heterogeneity also exists within Hip-Hop. In AotM, Hip-Hop also has the highest $B(C)$ among large communities (0.425), consistent with Silver et al.'s high-boundary prediction, a partial agreement not observed in AWA. The high $B(C)$ of dance-electronic in AWA contradicts Silver's prediction but may reflect the isolated consumption sphere of EDM listeners in the Japanese market.

High-$D(C)$ communities reveal how a single genre label conceals organized internal heterogeneity. C3 (rock-punk, $D=8$), the community with the highest $D(C)$, was decomposed into 8 significant subcommunities whose genre compositions differ qualitatively: pop-rock (pop + rock-punk mixed), rock-dominant (rock-punk 72--88\%), and R\&B fusion (rb-soul 48--61\%), reflecting segmentation within ``rock-punk'' by era, style, and adjacent genres. The H2 $\Delta H = 0.569$ confirms that this internal structure corresponds to sorting of genre labels. Hip-Hop shows the highest $D(C)$ in AotM ($D=7$) and high $D(C)$ in AWA ($D=3$), consistently deviating from Silver et al.'s single-centered prediction across both datasets. AWA C6's subcommunities reflect generational and stylistic differentiation (veteran/new generation/trap/underground), while AotM C3's seven subcommunities suggest even richer internal structure in the English-language context. Both align with Lizardo's hidden heterogeneity observation, suggesting that Hip-Hop's internal differentiation is a structural feature transcending cultural contexts and time periods.

The statistical relationship between $B(C)$ and external connection concentration (H3) can be illustrated at the individual community level. C7 (classical) is nearly completely isolated, and 79\% of C3's (rock-punk) external connections are concentrated in the other rock-punk community (C1). In contrast, C0 (pop) connects to 39 communities with a maximum of only 24\%, functioning as a mainstream hub. Among C6's (hiphop) external adjacent artists, only 3\% carry the hiphop genre label, indicating that boundary artists exhibit cross-genre consumption patterns leaning toward R\&B/pop.

\subsection{RQ3: Discovery beyond labels}

The bottom-up community extraction and two-dimensional framework reveal structures invisible to existing labels.

\subsubsection{Unnamed communities}

Multiple communities were detected that are strongly cohesive in consumption behavior but not fully described by existing genre labels.

\begin{table}
\centering
\scriptsize
\begin{tabular}{llrrp{55mm}}
\toprule
Community & Genre & $B(C)$ & $D(C)$ & Characteristics \\
\midrule
C5 & pop (73\%) & 0.735 & 2 & Independent consumption sphere with high pop purity. Labeled ``pop'' but behaviorally separated from mainstream pop (C0, C2, C4) \\
C16 & pop (54\%) & 0.984 & 0 & Virtually isolated with extremely high $B(C)$. Contains classical and soundtrack elements \\
C20 & rb-soul (44\%) & 0.782 & 0 & Consumption sphere where three labels (rb-soul, jazz-blues, and reggae-dub) merge \\
\bottomrule
\end{tabular}\end{table}

These are classified as ``pop'' or ``rb-soul'' under existing label systems, but they form independent consumption spheres structurally distinct from mainstream pop and rb-soul communities. Such structures can only be discovered through bottom-up methods.

\subsubsection{One label $\rightarrow$ multiple communities (label splitting)}

As a direct empirical demonstration of Lizardo~\cite{Lizardo2024}'s observation that ``broad labels conceal hidden heterogeneity,'' in AWA the ``pop'' label is distributed across 12 communities. Particularly structurally interesting is the splitting of rock-punk:

\begin{table}
\centering
\caption{Rock-punk label splitting in AWA.}
\begin{tabular}{llllll}
\toprule
Community & Size & $B(C)$ & $D(C)$ & Ext.\ communities & Top ext.\ share \\
\midrule
C1 & 423 & 0.516 & 4 & 17 & 0.37 \\
C3 & 302 & 0.821 & 8 & 7 & 0.79 \\
\bottomrule
\end{tabular}\end{table}

Even within the same ``rock-punk'' label, C1 is a low-$B(C)$ community with external connections dispersed across 17 communities, while C3 is a high-$B(C)$ community where 79\% of external connections are concentrated in a single community (C1). The difference in $B(C)$ captures qualitative differences in consumption spheres: ``pop-leaning'' with weak boundaries versus ``core rock'' with closed boundaries.

In AotM, Rock/Pop similarly splits into multiple communities (C0, C2, C3, C4), confirming that broad label splitting is reproduced across cultural contexts.

\subsubsection{Multiple labels $\rightarrow$ one community (label merging)}

Cases where labels designate separate genres but consumption behavior aggregates them into the same cluster.

C0 (679 artists) is the largest community, merging three major labels: pop (29\%), rb-soul (24\%), and dance-electronic (20\%). Its genre entropy is the highest of all communities (2.58), but $B(C) = 0.695$ is moderate, functioning as the ``boundaryless center'' of the mainstream pop sphere.

AotM's C1 (Indie Rock, 1,049 artists) is the second-largest community but has $D(C) = 1$, with minimal internal differentiation, functioning largely as a single consumption sphere in playlist terms.

\subsubsection{High $B(C)$ but low label purity}

Communities that appear ``mixed'' according to existing labels but are strongly closed behaviorally represent the most direct demonstration of the limitations of label systems.

C7 (63 artists) has four labels mixed (classical 30\%, pop 25\%, soundtrack 17\%, jazz-blues 14\%), yet exhibits the highest $B(C)$ (0.988) of all 29 communities. Its external adjacency is limited to only 2 nodes, making it virtually isolated.

In the H2 analysis, this community showed the largest $\Delta H = 0.778$ of all communities. Subcommunity decomposition at $D(C) = 3$ segmented the community into a classical cluster, a contemporary classical/ambient cluster, and a jazz cluster, organizing the label-level mixture into organized internal structure.

This case demonstrates that high $B(C)$ and label purity are independent. ``Mixed labels $\neq$ weak boundary''; listeners consume artists spanning multiple labels as a single closed consumption sphere.

\subsubsection{Non-LCC components: Isolated consumption spheres}

In AWA, 36.5\% of the network (1,595 nodes across 684 components) lies outside the LCC. These components have zero connections to the LCC, representing complete structural separation from the mainstream. Their genre composition differs markedly from the LCC: soundtrack (14.2\%) and dance-electronic (15.0\%) are substantially overrepresented. Soundtrack, in particular, forms multiple pure isolated components (7+ components with $>$50\% soundtrack) in addition to the near-isolated LCC community C15 ($B(C) = 0.986$). Jazz-blues forms no independent community within the LCC (it appears only as a sub-element of other communities) but does form independent components outside the LCC, consistent with Silver et al.~\cite{Silver2016}'s report of Jazz's extremely low crossing rate (2.1\%). A detailed breakdown of non-LCC components by genre is provided in Appendix~\ref{app:nonlcc}.

These non-LCC components can be interpreted as the endpoint of a continuous spectrum of external closure: from high-$B(C)$ communities within the LCC (C7: $B(C) = 0.988$, 2 external neighbors) to completely disconnected components (0 external neighbors). The existence of these isolated consumption spheres suggests that the music consumption space is not a single continuum but a heterogeneous structure containing multiple listening spheres structurally detached from the mainstream.

% ===================================================================
% 6. DISCUSSION
% ===================================================================
\section{Discussion}

\subsection{Theoretical implications of the two-dimensional framework}

The independence of $B(C)$ and $D(C)$ after controlling for community size, confirmed across two datasets from different platforms, cultural contexts, and time periods (AWA: partial $\rho = -0.080$, $p = 0.685$; AotM: partial $\rho = -0.361$, $p = 0.248$), is not merely a statistical convenience for constructing a two-dimensional coordinate system. It reflects a deeper theoretical regularity that multiple traditions in cultural sociology have anticipated from different angles.

DiMaggio~\cite{DiMaggio1987} proposed that classification systems in the arts vary along independent dimensions of \emph{ritual strength} (the intensity with which genre boundaries are defended) and \emph{differentiation} (the number of subdivisions). He argued that these dimensions are driven by distinct social processes: ritual strength by taste as a form of social identification and institutional enforcement, differentiation by competition and distinction-seeking among producers. Our $B(C)$ and $D(C)$ operationalize these concepts, and their empirical independence confirms DiMaggio's theoretical premise.

Bourdieu~\cite{Bourdieu1993}'s field theory offers a structural explanation for this independence. The \emph{autonomy} of a cultural field (its insulation from external economic and political forces) determines boundary strength, while the \emph{internal position structure} (the diversity of positions occupied by competing actors) determines differentiation. These are governed by different mechanisms: autonomy by the field's relationship to external power structures, internal differentiation by within-field strategies of distinction. A highly autonomous field can be internally homogeneous or diverse; conversely, a field with porous boundaries can still harbor rich internal structure.

Lena and Peterson~\cite{LenaPeterson2008}'s genre lifecycle model provides dynamic evidence. Their four genre types, Avant-garde (low boundary, low differentiation), Scene-based (moderate boundary, moderate differentiation), Industry-based (high boundary, high differentiation), and Traditionalist (high boundary, low differentiation), demonstrate that boundary strength and internal differentiation change \emph{asynchronously} across lifecycle stages. The four quadrants we observe in the $B(C) \times D(C)$ plane may thus represent communities at different developmental stages coexisting within the same network snapshot.

By operationalizing boundary strength and internal differentiation as Conductance Deficit and Wilcoxon-tested recursive decomposition, respectively, we enable continuous quantitative measurement of these theoretically independent dimensions. The cross-platform replication demonstrates that this independence is not an artifact of a specific platform or cultural context, but a general property of music community structure observable in consumption behavior data.

\subsection{Visualizing the limitations of existing label systems}

The case studies demonstrate the limitations of existing genre labels from three directions.

First, a single label splits into multiple communities with different $B(C)$ values (Section~5.4). The two rock-punk communities have similar $D(C)$ but markedly different $B(C)$, reflecting qualitative differences as consumption spheres.

Second, multiple labels merge into a single high-$B(C)$ community (Section~5.4). C7 exhibits the highest $B(C)$ despite containing four mixed labels, contradicting the intuition that ``mixed labels = weak boundary.''

Third, consumption spheres exist that are undefined by the label system (Section~5.4). These ``unnamed communities'' are structures discoverable only through bottom-up methods.

These findings serve as concrete empirical demonstrations of Lizardo~\cite{Lizardo2024}'s ``two problems concealed by macro-genre labels'': boundary overlap and internal heterogeneity.

\subsection{Agreement and disagreement with prior research}

In comparing with Silver et al.'s predictions, the multi-centered structure (high $D(C)$) of Rock and the low $B(C)$ of Pop were consistently confirmed. In contrast, Hip-Hop shows high $D(C)$ in both datasets, consistently deviating from Silver et al.'s single-centered prediction. This deviation aligns with Lizardo~\cite{Lizardo2024}'s observation that hidden heterogeneity also exists within Hip-Hop.

Differences in data sources may generate this discrepancy. Silver et al.'s data consists of musicians' \emph{self-reported genres}, reflecting genre affiliation as identity. Our data, by contrast, consists of \emph{consumers' playlist co-occurrences}, reflecting genre consumption as listening behavior. The homogeneity with which Hip-Hop musicians declare themselves ``Hip-Hop'' and the diversity with which consumers listen to Hip-Hop alongside other genres may capture different facets of the same phenomenon.

\subsection{Non-LCC components and heterogeneity of the music consumption space}

As shown in Section~5.4, 36.5\% of the AWA network's nodes exist outside the LCC, containing multiple isolated consumption spheres for soundtrack, jazz-blues, dance-electronic, and other genres. This suggests that the music consumption space is not a single continuum centered on the mainstream, but a heterogeneous space containing multiple consumption spheres that are structurally separated.

Soundtrack, in particular, consistently exhibits extremely high external closure from within the LCC (C15, $B(C)=0.986$) to non-LCC components (complete isolation). While prior research has not directly discussed the boundary strength of soundtrack, our results provide novel empirical evidence that listening behavior functionally tied to media consumption (film, anime, etc.) generates the strongest boundary structures among music genres.

The finding that jazz-blues does not form an independent community within the LCC and appears only in non-LCC components is also consistent with Silver et al.~\cite{Silver2016}'s report that Jazz's crossing rate is 2.1\% (among the lowest of all genres). The finding that Jazz's consumption sphere is structurally isolated from the mainstream is theoretically significant.

\subsection{Limitations}

Several limitations should be noted. First, the absolute value of the Conductance Deficit depends on network density, so AWA and AotM CD values cannot be directly compared; the high/low classification of $B(C)$ is based on relative rank order within each dataset, and cross-dataset comparison of boundary strength quality is out of scope. Second, $D(C)$ has a strong positive correlation with community size (AWA: $\rho = 0.782$; AotM: $\rho = 0.795$). While it is structurally natural that larger communities are more likely to contain subcommunities, when a small community is classified as single-centered, it is difficult to distinguish whether it is genuinely homogeneous or lacking detection power. The case of AotM's Indie Rock (C1, 1,049 artists) with $D(C) = 1$, minimal internal differentiation despite being the second-largest community, serves as a counterexample showing that size alone does not determine $D(C)$. Third, the Leiden algorithm assigns each artist to only one community, and thus cannot represent overlapping memberships across genres. As Ahn et al.~\cite{Ahn2010} showed, network communities can inherently possess both overlap and hierarchy, but this study focuses on tree-like internal structure. Fourth, both datasets are static snapshots from 2023 (AWA) and 1998--2011 (AotM); dynamic changes in genre structure remain unexamined. Fifth, to enable cross-platform comparison, AotM playlists were truncated to the first 8 tracks and the co-occurrence threshold was calibrated to match AWA's network density. While this normalization substantially improved the resolution of community structure (from 8 to 17 communities), the truncation discards information from later tracks in longer playlists. AotM also lacks artist-level genre labels, limiting label-dependent analyses to AWA (details and sensitivity analyses are reported in Appendix~\ref{app:preprocessing}). Finally, the first-level partition is seed-independent through consensus clustering, but subcommunity decomposition relies on a single seed (seed=42). Silver et al.~\cite{Silver2016} also applied a deterministic algorithm once, and we follow this precedent, but sensitivity analysis with multiple seeds remains a future task.

% ===================================================================
% 7. CONCLUSION
% ===================================================================
\section{Conclusion}

This study proposed and validated a quantitative framework for describing the community structure of playlist co-occurrence networks along two axes: external closure $B(C)$ and internal differentiation $D(C)$. The findings are organized around the three research questions.

\textbf{RQ1 (Framework validity):} We defined Conductance Deficit ($B(C)$) and recursive subcommunity decomposition ($D(C)$) as operationalizations of Silver et al.~\cite{Silver2016}'s theoretical framework and confirmed the validity of the two-dimensional coordinate system through three statistical hypotheses. Critically, the key results were replicated across two independent datasets from different platforms, cultural contexts, and time periods: $B(C)$ and $D(C)$ are independent after size control in both datasets (H1: AWA partial $\rho = -0.080$; AotM partial $\rho = -0.361$, both non-significant); $D(C)$ correlates with entropy reduction of existing labels (H2: $\rho = 0.518$**, AWA only due to label availability); and $B(C)$ correlates with concentration of external connections in both datasets (H3: AWA $\rho = -0.612$***; AotM $\rho = -0.853$***).

\textbf{RQ2 (Theoretical correspondence):} The observed $B \times D$ landscape is broadly consistent with Silver et al.~\cite{Silver2016}'s predictions, but Hip-Hop's internal differentiation consistently deviates from the single-centered prediction across both datasets (AWA: $D = 3$; AotM: $D = 7$), empirically supporting Lizardo~\cite{Lizardo2024}'s observation of hidden heterogeneity.

\textbf{RQ3 (Discovery beyond labels):} The bottom-up approach revealed structures invisible to existing labels: a single label splitting into multiple communities with different $B(C)$ values, multiple labels merging into a high-$B(C)$ community, and the existence of consumption spheres undefined by the label system. Furthermore, 36.5\% of the network exists outside the LCC as isolated consumption spheres, with specific genres such as soundtrack and jazz-blues being structurally separated from the mainstream.

Future work includes (1) dynamic analysis of genre structure using time-series data: the $B(C) \times D(C)$ coordinate system established here provides a natural basis for tracking how individual communities' boundary strength and internal differentiation evolve over time; (2) application of community detection methods that allow overlap; (3) development of normalization methods to mitigate the size dependence of $D(C)$; and (4) extension to additional platforms (e.g., Spotify, Apple Music) to further test generalizability beyond the two platforms examined here.

\backmatter

%\bmhead{Supplementary information}

%If your article has accompanying supplementary file/s please state so here. 

%Authors reporting data from electrophoretic gels and blots should supply the full unprocessed scans for key as part of their Supplementary information. This may be requested by the editorial team/s if it is missing.

%Please refer to Journal-level guidance for any specific requirements.

%\bmhead{Acknowledgements}

%Acknowledgements are not compulsory. Where included they should be brief. Grant or contribution numbers may be acknowledged.

%Please refer to Journal-level guidance for any specific requirements.

% ===================================================================
% LIST OF ABBREVIATIONS
% ===================================================================
\section*{List of abbreviations}

\begin{tabular}{@{}ll@{}}
LCC & Largest connected component \\
CD & Conductance Deficit \\
MIR & Music Information Retrieval \\
MSD & Million Song Dataset \\
NMI & Normalized Mutual Information \\
BH & Benjamini--Hochberg (false discovery rate correction) \\
EDM & Electronic dance music \\
\end{tabular}

% ===================================================================
% DECLARATIONS
% ===================================================================
\section*{Declarations}

\subsection*{Ethics approval and consent to participate}
Not applicable.

\subsection*{Consent for publication}
Not applicable.

\subsection*{Availability of data and materials}
The AWA data were obtained from a commercial music streaming service, and the raw data cannot be made publicly available. An anonymized network dataset with artist identifiers masked and genre labels attached is available from the corresponding author upon reasonable request for academic, non-commercial research purposes. The AotM-2011 dataset is publicly available from McFee and Lanckriet~\cite{McFee2012} (\url{https://brianmcfee.net/data/aotm2011.html}). Analysis code is planned for release on a GitHub repository upon paper acceptance.

\subsection*{Competing interests}
The author is employed by the company that operates the AWA music streaming service, from which one of the two datasets used in this study was obtained. There are no patents to declare.

\subsection*{Funding}
Not applicable.

\subsection*{Authors' contributions}
The author conceived and designed the study, performed all data analysis, developed the proposed framework, and wrote the manuscript.

\subsection*{Acknowledgements}
The author thanks Yukie Sano for valuable discussions on this research.

%%===================================================%%
%% For presentation purpose, we have included        %%
%% \bigskip command. Please ignore this.             %%
%%===================================================%%
%\bigskip
%\begin{flushleft}%
%Editorial Policies for:

%\bigskip\noindent
%Springer journals and proceedings: \url{https://www.springer.com/gp/editorial-policies}

%\bigskip\noindent
%Nature Portfolio journals: \url{https://www.nature.com/nature-research/editorial-policies}

%\bigskip\noindent
%\textit{Scientific Reports}: \url{https://www.nature.com/srep/journal-policies/editorial-policies}

%\bigskip\noindent
%BMC journals: \url{https://www.biomedcentral.com/getpublished/editorial-policies}
%\end{flushleft}

% ===================================================================
% APPENDIX
% ===================================================================
\begin{appendices}

\section{Cross-Platform Preprocessing and Threshold Calibration}
\label{app:preprocessing}

\subsection{Problem identification}

The AotM-2011 dataset has substantially longer playlists than AWA (median 19 vs.\ 8 tracks). Because the number of co-occurrence pairs per playlist scales quadratically with playlist length ($\binom{n}{2}$), AotM playlists generate 5.9 times more pairs on average (165 vs.\ 28). Without correction, this produces a network that is excessively dense (density 0.020 vs.\ AWA 0.006), yielding low modularity ($Q = 0.27$) and few communities (8 total, 6 with size $\geq 5$).

\subsection{Root cause analysis}

We identified two factors contributing to the density difference:

\textbf{Playlist length (primary factor).} Truncating AotM playlists to the first 8 tracks reduces the number of co-occurrence edges by 84--95\% across all thresholds.

\textbf{Artist population size (secondary factor).} AotM contains 103,801 unique artists (after truncation) compared to AWA's 21,665. Under a random null model, the expected co-occurrence count per pair is $\lambda = n_{\text{playlists}} \times k(k-1) / (N(N-1))$:
\begin{itemize}
    \item AWA: $\lambda = 246{,}110 \times 56 / (21{,}665 \times 21{,}664) = 0.029$
    \item AotM (truncated): $\lambda = 101{,}343 \times 56 / (103{,}801 \times 103{,}800) = 0.0005$
\end{itemize}

The 56-fold difference means that even identical thresholds represent very different filtering strengths.

\subsection{Two-step normalization procedure}

\textbf{Step 1: Playlist-length normalization (Trunc8).} Each AotM playlist is truncated to its first 8 tracks, matching AWA's playlist length. AotM playlists are user-curated mixtapes with intentional track ordering; the first-8 selection is deterministic and reproducible.

\textbf{Step 2: Density-matched threshold calibration.} We select the co-occurrence threshold that produces a network density closest to AWA's (0.006).

\begin{table}
\centering
\caption{AotM Trunc8 threshold calibration.}
\begin{tabular}{lllll}
\toprule
AotM Trunc8 threshold & LCC nodes & LCC edges & Density & Density ratio (vs.\ AWA 0.006) \\
\midrule
3 & 7,437 & 125,265 & 0.0045 & $0.71\times$ \\
4 & 4,671 & 75,948 & 0.0070 & $\mathbf{1.11\times}$ \\
5 & 3,392 & 52,429 & 0.0091 & $1.44\times$ \\
10 & 1,445 & 16,942 & 0.0163 & $2.6\times$ \\
20 & 632 & 5,152 & 0.0258 & $4.1\times$ \\
\bottomrule
\end{tabular}\end{table}

Threshold = 4 yields density 0.007, closest to AWA's 0.006 (ratio 1.11).

\subsection{Effect of normalization}

\begin{table}
\centering
\scriptsize
\begin{tabular}{lrrrr}
\toprule
 & Original ($\geq$20) & Trunc8 ($\geq$20) & Trunc8 ($\geq$4) & AWA ($\geq$50) \\
\midrule
Normalization & None & Step 1 & Step 1+2 & Ref. \\
LCC nodes & 2,606 & 632 & 4,671 & 2,779 \\
Density & 0.020 & 0.026 & 0.007 & 0.006 \\
Mean degree & 51.1 & 16.3 & 32.5 & 17.5 \\
Modularity & 0.273 & 0.325 & 0.308 & 0.715 \\
Communities & 8 & 6 & 17 & 65 \\
Size $\geq 5$ & 6 & 6 & 13 & 29 \\
\bottomrule
\end{tabular}
\caption{Effect of normalization on AotM network properties.}
\end{table}

\subsection{AotM original (without normalization): threshold sensitivity}

\begin{table}
\centering
\caption{AotM original threshold sensitivity.}
\begin{tabular}{lllll}
\toprule
Threshold & LCC nodes & LCC edges & Density & Mean degree \\
\midrule
5 & 10,974 & 435,003 & 0.007 & 79.3 \\
10 & 5,144 & 171,452 & 0.013 & 66.7 \\
20 & 2,606 & 66,633 & 0.020 & 51.1 \\
50 & 1,081 & 17,264 & 0.030 & 31.9 \\
100 & 519 & 5,319 & 0.040 & 20.5 \\
\bottomrule
\end{tabular}\end{table}

Lowering the threshold without playlist-length normalization increases LCC size but does not resolve the structural issue: at threshold = 5, the top 5 communities account for 98\% of all nodes, and macro/micro effect direction reverses relative to AWA.

% -------------------------------------------------------------------

\section{Non-LCC Component Details}
\label{app:nonlcc}

\subsection{Non-LCC component summary (AWA)}

After applying the co-occurrence threshold ($\geq 50$) to the AWA network (4,374 nodes), 685 connected components were identified. The LCC comprises 2,779 nodes (63.5\%). The remaining 1,595 nodes (36.5\%) are distributed across 684 non-LCC components. Most are micro-components of size 2--3 (648 components), but 19 components of size $\geq 5$ exist.

\subsection{Non-LCC components of size $\geq 5$ by genre}

\begin{table}
\centering
\caption{Non-LCC components of size $\geq 5$ (AWA).}
\small
\begin{tabular}{lllll}
\toprule
Component & Size & Primary genre composition & Density & Characteristics \\
\midrule
Comp1 & 25 & soundtrack 72\%, pop 20\% & 0.113 & Film music composers \\
Comp2 & 19 & soundtrack 94\% & 0.433 & Nearly pure soundtrack sphere \\
Comp3 & 12 & animation-vocaloid 92\% & 0.773 & Anime/Vocaloid sphere \\
Comp5 & 9 & jazz-blues 67\%, rb-soul 22\% & 0.222 & Jazz isolated sphere \\
Comp6 & 8 & reggae-dub 71\% & 0.500 & Reggae isolated sphere \\
Comp8 & 7 & dance-electronic 100\% & 1.000 & Fully connected EDM cluster \\
Comp13 & 5 & rock-punk 60\%, jazz-blues 40\% & 0.600 & Fusion-type sphere \\
\bottomrule
\end{tabular}\end{table}

\subsection{Genre distribution of non-LCC components}

The genre distribution across all non-LCC nodes with labels (1,555 nodes): pop (27.0\%), dance-electronic (15.0\%), rb-soul (14.6\%), soundtrack (14.2\%), rock-punk (10.7\%). Compared to the LCC, soundtrack and dance-electronic are substantially overrepresented.

\subsection{Soundtrack structural isolation}

At least 7 non-LCC components have soundtrack proportions exceeding 50\%. Combined with the LCC-internal soundtrack community (C15, size = 14, $B(C) = 0.986$), soundtrack consumption occurs in multiple closed listening spheres structurally separated from the mainstream. This pattern is consistent with listening behavior functionally tied to media consumption (film, anime, games) being unlikely to generate cross-genre co-occurrence edges.

\subsection{Jazz-blues structural isolation}

No community with jazz-blues as its primary genre exists within the LCC; it appears only as a sub-element of the classical community (C7) and the rb-soul community (C20). In contrast, non-LCC components include jazz-blues-dominant components (Comp5, size = 9) and fusion-type components (Comp13, size = 5). The jazz consumption sphere has very weak connections to the mainstream, appearing as an independent consumption sphere only outside the LCC.

\end{appendices}

%%===========================================================================================%%
%% If you are submitting to one of the Nature Portfolio journals, using the eJP submission   %%
%% system, please include the references within the manuscript file itself. You may do this  %%
%% by copying the reference list from your .bbl file, paste it into the main manuscript .tex %%
%% file, and delete the associated \verb+\bibliography+ commands.                            %%
%%===========================================================================================%%

\bibliography{sn-bibliography}% common bib file

%% BioMed_Central_Bib_Style_v1.01

\begin{thebibliography}{26}
% BibTex style file: bmc-mathphys.bst (version 2.1), 2014-07-24
\ifx \bisbn   \undefined \def \bisbn  #1{ISBN #1}\fi
\ifx \binits  \undefined \def \binits#1{#1}\fi
\ifx \bauthor  \undefined \def \bauthor#1{#1}\fi
\ifx \batitle  \undefined \def \batitle#1{#1}\fi
\ifx \bjtitle  \undefined \def \bjtitle#1{#1}\fi
\ifx \bvolume  \undefined \def \bvolume#1{\textbf{#1}}\fi
\ifx \byear  \undefined \def \byear#1{#1}\fi
\ifx \bissue  \undefined \def \bissue#1{#1}\fi
\ifx \bfpage  \undefined \def \bfpage#1{#1}\fi
\ifx \blpage  \undefined \def \blpage #1{#1}\fi
\ifx \burl  \undefined \def \burl#1{\textsf{#1}}\fi
\ifx \doiurl  \undefined \def \doiurl#1{\url{https://doi.org/#1}}\fi
\ifx \betal  \undefined \def \betal{\textit{et al.}}\fi
\ifx \binstitute  \undefined \def \binstitute#1{#1}\fi
\ifx \binstitutionaled  \undefined \def \binstitutionaled#1{#1}\fi
\ifx \bctitle  \undefined \def \bctitle#1{#1}\fi
\ifx \beditor  \undefined \def \beditor#1{#1}\fi
\ifx \bpublisher  \undefined \def \bpublisher#1{#1}\fi
\ifx \bbtitle  \undefined \def \bbtitle#1{#1}\fi
\ifx \bedition  \undefined \def \bedition#1{#1}\fi
\ifx \bseriesno  \undefined \def \bseriesno#1{#1}\fi
\ifx \blocation  \undefined \def \blocation#1{#1}\fi
\ifx \bsertitle  \undefined \def \bsertitle#1{#1}\fi
\ifx \bsnm \undefined \def \bsnm#1{#1}\fi
\ifx \bsuffix \undefined \def \bsuffix#1{#1}\fi
\ifx \bparticle \undefined \def \bparticle#1{#1}\fi
\ifx \barticle \undefined \def \barticle#1{#1}\fi
\bibcommenthead
\ifx \bconfdate \undefined \def \bconfdate #1{#1}\fi
\ifx \botherref \undefined \def \botherref #1{#1}\fi
\ifx \url \undefined \def \url#1{\textsf{#1}}\fi
\ifx \bchapter \undefined \def \bchapter#1{#1}\fi
\ifx \bbook \undefined \def \bbook#1{#1}\fi
\ifx \bcomment \undefined \def \bcomment#1{#1}\fi
\ifx \oauthor \undefined \def \oauthor#1{#1}\fi
\ifx \citeauthoryear \undefined \def \citeauthoryear#1{#1}\fi
\ifx \endbibitem  \undefined \def \endbibitem {}\fi
\ifx \bconflocation  \undefined \def \bconflocation#1{#1}\fi
\ifx \arxivurl  \undefined \def \arxivurl#1{\textsf{#1}}\fi
\csname PreBibitemsHook\endcsname

%%% 1
\bibitem[\protect\citeauthoryear{Lizardo}{2024}]{Lizardo2024}
\begin{barticle}
\bauthor{\bsnm{Lizardo}, \binits{O.}}:
\batitle{From macrogenres to microgenres via relationality}.
\bjtitle{Poetics}
\bvolume{102},
\bfpage{101868}
(\byear{2024})
\doiurl{10.1016/j.poetic.2023.101868}
\end{barticle}
\endbibitem

%%% 2
\bibitem[\protect\citeauthoryear{Silver et~al.}{2016}]{Silver2016}
\begin{barticle}
\bauthor{\bsnm{Silver}, \binits{D.}},
\bauthor{\bsnm{Lee}, \binits{M.}},
\bauthor{\bsnm{Childress}, \binits{C.C.}}:
\batitle{Genre complexes in popular music}.
\bjtitle{PLOS ONE}
\bvolume{11}(\bissue{5}),
\bfpage{0155471}
(\byear{2016})
\doiurl{10.1371/journal.pone.0155471}
\end{barticle}
\endbibitem

%%% 3
\bibitem[\protect\citeauthoryear{Schedl et~al.}{2014}]{SchedlGomez2014}
\begin{barticle}
\bauthor{\bsnm{Schedl}, \binits{M.}},
\bauthor{\bsnm{G{\'o}mez}, \binits{E.}},
\bauthor{\bsnm{Urbano}, \binits{J.}}:
\batitle{Music information retrieval: Recent developments and applications}.
\bjtitle{Foundations and Trends in Information Retrieval}
\bvolume{8}(\bissue{2--3}),
\bfpage{127}--\blpage{261}
(\byear{2014})
\doiurl{10.1561/1500000042}
\end{barticle}
\endbibitem

%%% 4
\bibitem[\protect\citeauthoryear{Bogdanov et~al.}{2019}]{Bogdanov2019}
\begin{bchapter}
\bauthor{\bsnm{Bogdanov}, \binits{D.}},
\bauthor{\bsnm{Porter}, \binits{A.}},
\bauthor{\bsnm{Schreiber}, \binits{H.}},
\bauthor{\bsnm{Urbano}, \binits{J.}},
\bauthor{\bsnm{Oramas}, \binits{S.}}:
\bctitle{The {AcousticBrainz} genre dataset: Multi-source, multi-level, multi-label, and large-scale}.
In: \bbtitle{Proceedings of the 20th International Society for Music Information Retrieval Conference}.
\bsertitle{ISMIR 2019},
pp. \bfpage{206}--\blpage{213}
(\byear{2019}).
\burl{https://mtg.github.io/acousticbrainz-genre-dataset/}
\end{bchapter}
\endbibitem

%%% 5
\bibitem[\protect\citeauthoryear{DiMaggio}{1987}]{DiMaggio1987}
\begin{barticle}
\bauthor{\bsnm{DiMaggio}, \binits{P.}}:
\batitle{Classification in art}.
\bjtitle{American Sociological Review}
\bvolume{52}(\bissue{4}),
\bfpage{440}--\blpage{455}
(\byear{1987})
\end{barticle}
\endbibitem

%%% 6
\bibitem[\protect\citeauthoryear{Kov{\'a}cs and Hannan}{2015}]{KovacsHannan2015}
\begin{barticle}
\bauthor{\bsnm{Kov{\'a}cs}, \binits{B.}},
\bauthor{\bsnm{Hannan}, \binits{M.T.}}:
\batitle{Conceptual spaces and the consequences of category spanning}.
\bjtitle{Sociological Science}
\bvolume{2},
\bfpage{252}--\blpage{286}
(\byear{2015})
\doiurl{10.15195/v2.a13}
\end{barticle}
\endbibitem

%%% 7
\bibitem[\protect\citeauthoryear{Shi et~al.}{2018}]{Shi2018}
\begin{barticle}
\bauthor{\bsnm{Shi}, \binits{Y.}},
\bauthor{\bsnm{Lim}, \binits{Y.}},
\bauthor{\bsnm{Suh}, \binits{C.S.}}:
\batitle{Innovation or deviation? the relationship between boundary crossing and audience evaluation in the music field}.
\bjtitle{PLOS ONE}
\bvolume{13}(\bissue{10}),
\bfpage{0203065}
(\byear{2018})
\doiurl{10.1371/journal.pone.0203065}
\end{barticle}
\endbibitem

%%% 8
\bibitem[\protect\citeauthoryear{Schreiber}{2015}]{Schreiber2015}
\begin{bchapter}
\bauthor{\bsnm{Schreiber}, \binits{H.}}:
\bctitle{Improving genre annotations for the {Million Song Dataset}}.
In: \bbtitle{Proceedings of the 16th International Society for Music Information Retrieval Conference}.
\bsertitle{ISMIR 2015},
pp. \bfpage{241}--\blpage{247},
\bconflocation{M{\'a}laga, Spain}
(\byear{2015}).
\burl{https://archives.ismir.net/ismir2015/paper/000102.pdf}
\end{bchapter}
\endbibitem

%%% 9
\bibitem[\protect\citeauthoryear{Lamere}{2008}]{Lamere2008}
\begin{barticle}
\bauthor{\bsnm{Lamere}, \binits{P.}}:
\batitle{Social tagging and music information retrieval}.
\bjtitle{Journal of New Music Research}
\bvolume{37}(\bissue{2}),
\bfpage{101}--\blpage{114}
(\byear{2008})
\doiurl{10.1080/09298210802479284}
\end{barticle}
\endbibitem

%%% 10
\bibitem[\protect\citeauthoryear{Celma}{2010}]{Celma2010}
\begin{bbook}
\bauthor{\bsnm{Celma}, \binits{O.}}:
\bbtitle{Music Recommendation and Discovery in the Long Tail}.
\bpublisher{Springer},
\blocation{Berlin}
(\byear{2010}).
\doiurl{10.1007/978-3-642-13287-2}
\end{bbook}
\endbibitem

%%% 11
\bibitem[\protect\citeauthoryear{Jiang et~al.}{2024}]{JiangSpotify2024}
\begin{bchapter}
\bauthor{\bsnm{Jiang}, \binits{J.}},
\bauthor{\bsnm{Ponnada}, \binits{A.}},
\bauthor{\bsnm{Li}, \binits{A.}},
\bauthor{\bsnm{Lacker}, \binits{B.}},
\bauthor{\bsnm{Way}, \binits{S.F.}}:
\bctitle{A genre-based analysis of new music streaming at scale}.
In: \bbtitle{Proceedings of the 16th ACM Web Science Conference}.
\bsertitle{WebSci '24}
(\byear{2024}).
\doiurl{10.1145/3614419.3644002}
\end{bchapter}
\endbibitem

%%% 12
\bibitem[\protect\citeauthoryear{Corr{\^{e}}a et~al.}{2011}]{CorreaDe2011}
\begin{bchapter}
\bauthor{\bsnm{Corr{\^{e}}a}, \binits{D.C.}},
\bauthor{\bsnm{Levada}, \binits{A.L.M.}},
\bauthor{\bsnm{Costa}, \binits{L.d.F.}}:
\bctitle{Finding community structure in music genres networks}.
In: \bbtitle{Proceedings of the 12th International Society for Music Information Retrieval Conference}.
\bsertitle{ISMIR 2011},
pp. \bfpage{447}--\blpage{452},
\bconflocation{Miami, FL, USA}
(\byear{2011}).
\burl{https://ismir2011.ismir.net/papers/PS3-16.pdf}
\end{bchapter}
\endbibitem

%%% 13
\bibitem[\protect\citeauthoryear{Jiang and Huynh}{2022}]{JiangHuynh2022}
\begin{barticle}
\bauthor{\bsnm{Jiang}, \binits{Z.}},
\bauthor{\bsnm{Huynh}, \binits{H.N.}}:
\batitle{Unveiling music genre structure through common-interest communities}.
\bjtitle{Social Network Analysis and Mining}
\bvolume{12},
\bfpage{35}
(\byear{2022})
\doiurl{10.1007/s13278-022-00863-2}
\end{barticle}
\endbibitem

%%% 14
\bibitem[\protect\citeauthoryear{Park and Park}{2025}]{ParkPark2025}
\begin{barticle}
\bauthor{\bsnm{Park}, \binits{D.}},
\bauthor{\bsnm{Park}, \binits{J.}}:
\batitle{Evolution of sample-based music authorship network}.
\bjtitle{EPJ Data Science}
\bvolume{14}(\bissue{1}),
\bfpage{5}
(\byear{2025})
\doiurl{10.1140/epjds/s13688-025-00524-2}
\end{barticle}
\endbibitem

%%% 15
\bibitem[\protect\citeauthoryear{Park et~al.}{2015}]{ParkBaeSchichPark2015}
\begin{barticle}
\bauthor{\bsnm{Park}, \binits{D.}},
\bauthor{\bsnm{Bae}, \binits{A.}},
\bauthor{\bsnm{Schich}, \binits{M.}},
\bauthor{\bsnm{Park}, \binits{J.}}:
\batitle{Topology and evolution of the network of {Western} classical music composers}.
\bjtitle{EPJ Data Science}
\bvolume{4},
\bfpage{2}
(\byear{2015})
\doiurl{10.1140/epjds/s13688-015-0039-z}
\end{barticle}
\endbibitem

%%% 16
\bibitem[\protect\citeauthoryear{McFee and Lanckriet}{2011}]{McFee2011}
\begin{bchapter}
\bauthor{\bsnm{McFee}, \binits{B.}},
\bauthor{\bsnm{Lanckriet}, \binits{G.R.G.}}:
\bctitle{The natural language of playlists}.
In: \bbtitle{Proceedings of the 12th International Society for Music Information Retrieval Conference}.
\bsertitle{ISMIR 2011},
pp. \bfpage{537}--\blpage{542},
\bconflocation{Miami, FL, USA}
(\byear{2011})
\end{bchapter}
\endbibitem

%%% 17
\bibitem[\protect\citeauthoryear{McFee and Lanckriet}{2012}]{McFee2012}
\begin{bchapter}
\bauthor{\bsnm{McFee}, \binits{B.}},
\bauthor{\bsnm{Lanckriet}, \binits{G.R.G.}}:
\bctitle{Hypergraph models of playlist dialects}.
In: \bbtitle{Proceedings of the 13th International Society for Music Information Retrieval Conference}.
\bsertitle{ISMIR 2012},
pp. \bfpage{343}--\blpage{348},
\bconflocation{Porto, Portugal}
(\byear{2012}).
\doiurl{10.5281/zenodo.1415618}
\end{bchapter}
\endbibitem

%%% 18
\bibitem[\protect\citeauthoryear{Fortunato and Barth{\'e}lemy}{2007}]{Fortunato2007}
\begin{barticle}
\bauthor{\bsnm{Fortunato}, \binits{S.}},
\bauthor{\bsnm{Barth{\'e}lemy}, \binits{M.}}:
\batitle{Resolution limit in community detection}.
\bjtitle{Proceedings of the National Academy of Sciences}
\bvolume{104}(\bissue{1}),
\bfpage{36}--\blpage{41}
(\byear{2007})
\doiurl{10.1073/pnas.0605965104}
\end{barticle}
\endbibitem

%%% 19
\bibitem[\protect\citeauthoryear{Arenas et~al.}{2008}]{Arenas2008}
\begin{barticle}
\bauthor{\bsnm{Arenas}, \binits{A.}},
\bauthor{\bsnm{Fern{\'a}ndez}, \binits{A.}},
\bauthor{\bsnm{G{\'o}mez}, \binits{S.}}:
\batitle{Analysis of the structure of complex networks at different resolution levels}.
\bjtitle{New Journal of Physics}
\bvolume{10},
\bfpage{053039}
(\byear{2008})
\doiurl{10.1088/1367-2630/10/5/053039}
\end{barticle}
\endbibitem

%%% 20
\bibitem[\protect\citeauthoryear{Leskovec et~al.}{2008}]{Leskovec2008}
\begin{bchapter}
\bauthor{\bsnm{Leskovec}, \binits{J.}},
\bauthor{\bsnm{Lang}, \binits{K.J.}},
\bauthor{\bsnm{Dasgupta}, \binits{A.}},
\bauthor{\bsnm{Mahoney}, \binits{M.W.}}:
\bctitle{Statistical properties of community structure in large social and information networks}.
In: \bbtitle{Proceedings of the 17th International Conference on World Wide Web}.
\bsertitle{WWW '08},
pp. \bfpage{695}--\blpage{704}.
\bpublisher{ACM},
\blocation{New York, NY, USA}
(\byear{2008}).
\doiurl{10.1145/1367497.1367591}
\end{bchapter}
\endbibitem

%%% 21
\bibitem[\protect\citeauthoryear{}{}]{AWA}
\begin{botherref}
{AWA} -- Music streaming service.
Accessed 28 Mar 2026.
\url{https://awa.fm/}
\end{botherref}
\endbibitem

%%% 22
\bibitem[\protect\citeauthoryear{Traag et~al.}{2019}]{Traag2019}
\begin{barticle}
\bauthor{\bsnm{Traag}, \binits{V.A.}},
\bauthor{\bsnm{Waltman}, \binits{L.}},
\bauthor{\bsnm{Eck}, \binits{N.J.}}:
\batitle{From {Louvain} to {Leiden}: Guaranteeing well-connected communities}.
\bjtitle{Scientific Reports}
\bvolume{9},
\bfpage{5233}
(\byear{2019})
\doiurl{10.1038/s41598-019-41695-z}
\end{barticle}
\endbibitem

%%% 23
\bibitem[\protect\citeauthoryear{Lancichinetti and Fortunato}{2012}]{Lancichinetti2012}
\begin{barticle}
\bauthor{\bsnm{Lancichinetti}, \binits{A.}},
\bauthor{\bsnm{Fortunato}, \binits{S.}}:
\batitle{Consensus clustering in complex networks}.
\bjtitle{Scientific Reports}
\bvolume{2},
\bfpage{336}
(\byear{2012})
\doiurl{10.1038/srep00336}
\end{barticle}
\endbibitem

%%% 24
\bibitem[\protect\citeauthoryear{Bourdieu}{1993}]{Bourdieu1993}
\begin{bbook}
\bauthor{\bsnm{Bourdieu}, \binits{P.}}:
\bbtitle{The Field of Cultural Production: Essays on Art and Literature}.
\bpublisher{Columbia University Press},
\blocation{New York}
(\byear{1993})
\end{bbook}
\endbibitem

%%% 25
\bibitem[\protect\citeauthoryear{Lena and Peterson}{2008}]{LenaPeterson2008}
\begin{barticle}
\bauthor{\bsnm{Lena}, \binits{J.C.}},
\bauthor{\bsnm{Peterson}, \binits{R.A.}}:
\batitle{Classification as culture: Types and trajectories of music genres}.
\bjtitle{American Sociological Review}
\bvolume{73}(\bissue{5}),
\bfpage{697}--\blpage{718}
(\byear{2008})
\doiurl{10.1177/000312240807300501}
\end{barticle}
\endbibitem

%%% 26
\bibitem[\protect\citeauthoryear{Ahn et~al.}{2010}]{Ahn2010}
\begin{barticle}
\bauthor{\bsnm{Ahn}, \binits{Y.-Y.}},
\bauthor{\bsnm{Bagrow}, \binits{J.P.}},
\bauthor{\bsnm{Lehmann}, \binits{S.}}:
\batitle{Link communities reveal multiscale complexity in networks}.
\bjtitle{Nature}
\bvolume{466},
\bfpage{761}--\blpage{764}
(\byear{2010})
\doiurl{10.1038/nature09182}
\end{barticle}
\endbibitem

\end{thebibliography}
%% if required, the content of .bbl file can be included here once bbl is generated
%%\input sn-article.bbl

\end{document}